\documentclass[aps,prl,preprint,numerical,amssymb,natbib,square,superscriptaddress,graphicx]{revtex4-1}

\usepackage{hyperref}
\usepackage{ulem}
\hypersetup{colorlinks,citecolor=blue,filecolor=blue,linkcolor=blue,urlcolor=blue}
\usepackage{color}

\usepackage{verbatim}
\usepackage{graphicx}
\usepackage[T1]{fontenc}

\newcommand{\nuc}[2]{$^{#1}$#2}
\newcommand{\fsh}{$f_{7/2}$}

\newcommand{\tpl}{2$^+$}
\newcommand{\fpl}{4$^+$}
\newcommand{\spl}{6$^+$}
\newcommand{\fp}{$fp$}

\newcommand{\fthm}{$\frac{15}{2}^-$}
\newcommand{\sthm}{$\frac{17}{2}^-$}
\newcommand{\nthm}{$\frac{19}{2}^-$}

\newcommand{\afiftytwo}{$^{52}$Ni$/^{52}$Cr}
\newcommand{\afiftyone}{$^{51}$Co$/^{51}$Cr}

\begin{document}

\preprint{to be submitted for publication in Physical Review Letters}

\title{Mirror energy differences at large isospin studied through direct two-nucleon knockout}


\author{P.J.\ Davies}
\email{paul.davies@york.ac.uk}
\author{M.A.\ Bentley}
\author{T.W.\ Henry}
\affiliation{Department of Physics, University of York, Heslington,
York YO10~5DD, UK}

\author{E.C.\ Simpson}
\affiliation{Department of Physics, Faculty of Engineering and Physical Sciences, University of Surrey, Guildford, GU2 7XH, UK}

\author{A.\ Gade}
\affiliation{National Superconducting Cyclotron Laboratory, Michigan State University, East Lansing, Michigan 48824, USA}

\author{S.M.\ Lenzi}
\affiliation{Dipartimento di Fisica del'Universit\`{a} and INFN, Sezione di Padova, I-35131 Padova, Italy}

\author{T.\ Baugher}
\author{D.\ Bazin}
\author{J.S.\ Berryman}
\affiliation{National Superconducting Cyclotron Laboratory, Michigan State University, East Lansing, Michigan 48824, USA}

\author{A.M.\ Bruce}
\affiliation{School of Computing, Engineering and Mathematics, University of Brighton, Brighton BN2 4GJ, United Kingdom}

\author{C.Aa.\ Diget}
\affiliation{Department of Physics, University of York, Heslington,
York YO10~5DD, UK}

\author{H.\ Iwasaki}
\author{A.\ Lemasson}
\author{S.\ McDaniel}
\affiliation{National Superconducting Cyclotron Laboratory, Michigan State University, East Lansing, Michigan 48824, USA}

\author{D.R.\ Napoli}
\affiliation{INFN, Laboratori Nazionali di Legnaro, I-35020 Legnaro, Italy}

\author{A.\ Ratkiewicz}
\affiliation{National Superconducting Cyclotron Laboratory, Michigan State University, East Lansing, Michigan 48824, USA}

\author{L.\ Scruton}
\affiliation{Department of Physics, University of York, Heslington,
York YO10~5DD, UK}

\author{A.\ Shore}
\author{R.\ Stroberg}
\affiliation{National Superconducting Cyclotron Laboratory, Michigan State University, East Lansing, Michigan 48824, USA}

\author{J.A.\ Tostevin}
\affiliation{Department of Physics, Faculty of Engineering and Physical Sciences, University of Surrey, Guildford, GU2 7XH, UK}

\author{D.\ Weisshaar}
\author{K.\ Wimmer}
\author{R.\ Winkler}
\affiliation{National Superconducting Cyclotron Laboratory, Michigan State University, East Lansing, Michigan 48824, USA}


\begin{abstract}
The first spectroscopy of excited states in \nuc{52}{Ni}~ ($T_z=-2$) and \nuc{51}{Co}~ ($T_z=-\frac{3}{2}$) has been obtained using the highly-selective two-neutron knockout reaction. Mirror energy differences between isobaric analogue states in these nuclei and their mirror partners are interpreted in terms of isospin non-conserving effects. Comparison between large-scale shell-model calculations and data provide the most compelling evidence to date that both electromagnetic and an additional isospin non-conserving interaction for $J=2$ couplings, of unknown origin, are required to obtain good agreement.
\end{abstract}

\date{\today}

\maketitle


Symmetries play a central role in physics, and can greatly simplify a model space and introduce experimentally verifiable predictions, such as conservation laws. In the case of the atomic nucleus the introduction of isospin ($T$), and the associated projection ($T_z=(N-Z)/2$) \cite{Heisenberg:1932iv}, the formalism by which the proton ($t_z=-\frac{1}{2}$) and neutron ($t_z=+\frac{1}{2}$) are treated as two states of the same particle, led Wigner to the concept of isospin symmetry \cite{Wigner:1937cl}. The symmetry is based on the assumption that nucleon-nucleon interactions are charge-symmetric and charge-independent. In the absence of isospin-breaking interactions, the model requires {\sl exact} symmetry (degeneracy) between analogue states in nuclei with the same mass but interchanged numbers of protons and neutrons (isobaric analogue states - IAS). Electromagnetic effects break the degeneracy, and provide an interaction that mixes states of different isospin. Historically, accounting for energy differences purely in terms of Coulomb effects has proved problematic (e.g. \cite{Nolen:td}), where predicted Coulomb Displacement Energies between IAS differed systematically from experimental values (the {\sl Nolen-Schiffer anomaly}), suggesting that other isospin-breaking effects need to be accounted for. Indeed, evidence of such symmetry breaking is found in the nucleon-nucleon interaction (e.g. \cite{Li:1998gp} and references therein). It has been suggested, e.g. \cite{Miller:1994tj,Shahnas:1994hv}, that this charge-symmetry breaking may contribute to the {\sl Nolen-Schiffer anomaly}, although the situation has not been fully resolved. 

Despite these difficulties, differences between {\sl excitation energies} of IAS (i.e. having normalised the ground state energies) show a remarkable degree of symmetry and can be well-reproduced by shell-model calculations which account for the electromagnetic multipole (i.e. two-body  Coulomb) effects \cite{Bentley:1998dh} and monopole effects \cite{Zuker:2002cf} affecting single-particle energies and related to radial and deformation changes. Detailed investigation of these phenomena has focused mainly on nuclei in the \fsh~shell, those nuclei between doubly magic \nuc{40}{Ca} and \nuc{56}{Ni} \cite{Bentley:2007fe,2005MPLA...20.2977E} although there are new results and shell-model calculations available in the $fpg$-shell also (e.g.~\cite{Kaneko}). The \fsh~region is particularly attractive for such an investigation, as $N<Z$ nuclei are experimentally accessible and large-scale shell-model calculations are known to provide an excellent description~\cite{Poves:2001fi}. A systematic study of these nuclei shows that a single shell-model prescription can be used to reproduce the differences in the excitation energy of mirror pairs (mirror energy differences -- MED) with excellent accuracy~\cite{Bentley:2007fe}. In this prescription,  isospin-breaking effects can be accounted for by the electromagnetic interaction. However, in order to achieve the best fit to the experimental MED, it is found \cite{Bentley:2007fe,Williams:2003km,Zuker:2002cf} that an additional repulsive two-body matrix element needs to be added for \fsh~protons coupled to $J=2$. Moreover, the additional $J=2$ term required is comparable to the Coulomb contribution.  This is the so-called $J=2$ anomaly, which has caused much interest, so far without a satisfactory explanation.

The motivation for the current work, focussed on the $T_z=-2$ nucleus \nuc{52}{Ni}, was two fold. Firstly, extending these studies to excited states of mirror nuclei with large differences in proton number and where we approach the limits of nuclear binding (i.e. near the proton drip-line) provides a stringent test of the models developed. Secondly, calculations indicated that the $J=2$ anomaly would play a significant role in {\sl both} the $J^\pi=2^+$ and \fpl~states in the $A=52$ pair but be absent for the \spl~state, giving a direct prediction that can be tested. In this Letter we report on the first excited states identified in the exotic nuclei \nuc{52}{Ni}~ ($T_z=-2$) and \nuc{51}{Co}~ ($T_z=-\frac{3}{2}$) -- the latter of which is expected to have excited states which are all above the calculated 88 keV proton separation energy \cite{Audi:2003jn}. We used, we believe for the first time, the approach of mirrored two-nucleon knockout to identify the IAS in the $T_z=\pm 2$ nuclei \afiftytwo~ and to determine the mirror energy differences. The states identified are the highest spin states yet observed in any nuclei with $T_z=-2$ or $-\frac{3}{2}$, demonstrating the power of the direct two-neutron knockout approach for populating intermediate-spin states in these proton-rich systems. The MED results are compared with state-of-the-art shell-model calculations, and crucial information on the $J=2$ effect is extracted.


\begin{figure}
\includegraphics[width=\columnwidth]{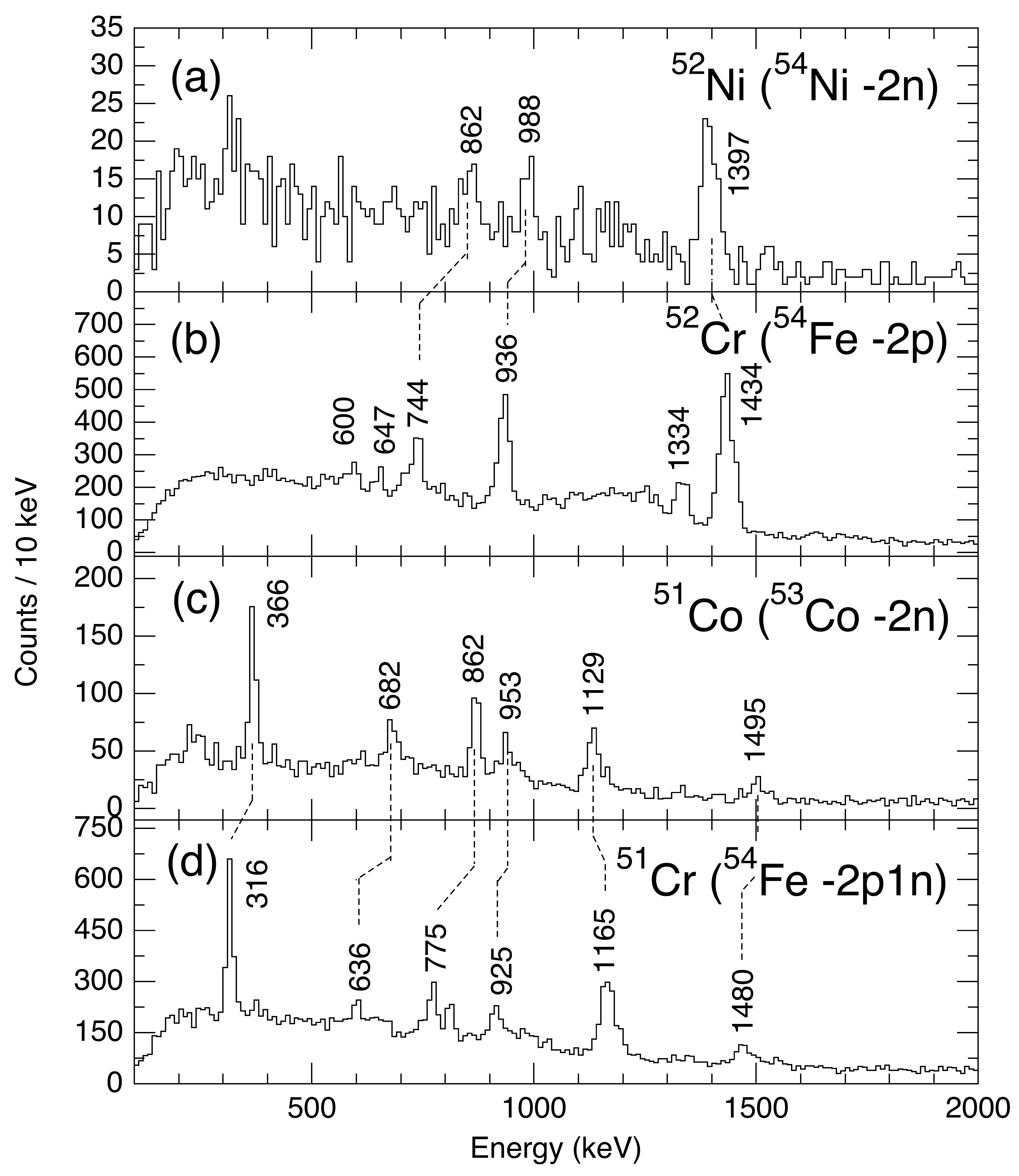}
\caption{$\gamma$-ray transitions found in coincidence with (a) \nuc{52}{Ni}~fragments, (b) \nuc{52}{Cr}, (c) \nuc{51}{Co}~and (d)\nuc{51}{Cr}. The dashed lines indicate proposed correspondence between the mirror nuclei.}
\label{spectra}
\end{figure}

The experiment was performed at the National Superconducting Cyclotron Laboratory at Michigan State University. The secondary beams of interest were produced by the fragmentation of a 160 AMeV beam of \nuc{58}{Ni} incident on a \nuc{9}{Be} primary target, with the resulting fragments separated by the A1900 fragment separator \cite{Morrissey:2003ex,Morrissey:1996ua} and identified downstream using their time-of-flight. 

The secondary target was located at the reaction target position of the S800 spectrograph \cite{Bazin:2003ex,Yurkon:1999ig}, and unique identification of the reaction products was achieved using the energy loss in an ion chamber and the time-of-flight through the S800. Surrounding the S800 reaction target position was SeGA, used to record $\gamma$ rays emitted in flight \cite{Mueller:2001ky}. 

Excited states in \nuc{52}{Ni} were populated by two-neutron knockout from~\nuc{54}{Ni}~ 
at 87 AMeV on the secondary \nuc{9}{Be} target of areal density 188 mg/cm$^2$. Since in this case the two nucleons are initially well bound, their sudden removal is expected to be direct as the indirect process (a single neutron removal followed by neutron evaporation) is not favoured energetically \cite{Tostevin:2006ev}. Two-neutron knockout was chosen here for two reasons. Firstly, the cross-section is expected to be large as both the \nuc{54}{Ni}~ and the yrast states of \nuc{52}{Ni}~ should be dominated by $\nu$(\fsh)  configurations. Secondly, the total angular momentum of the removed nucleons can be between $J=0$ and $J=6$ and hence, states up to intermediate spin should be populated. The presence of \nuc{53}{Co} in the secondary-beam cocktail also allowed the study of excited states in \nuc{51}{Co}, for the first time. The mirror partner to \nuc{52}{Ni}~(\nuc{52}{Cr}) was populated through the mirrored reaction - two-proton knockout from the \nuc{54}{Fe} beam and the new states in \nuc{52}{Ni}~identified through the spectral comparison from the mirrored reactions (see e.g.~\cite{Brown:2009fc}). Finally, the mirror of \nuc{51}{Co}~(\nuc{51}{Cr}) was produced through 2p1n removal from \nuc{54}{Fe}.

\begin{figure}
\includegraphics[width=\columnwidth]{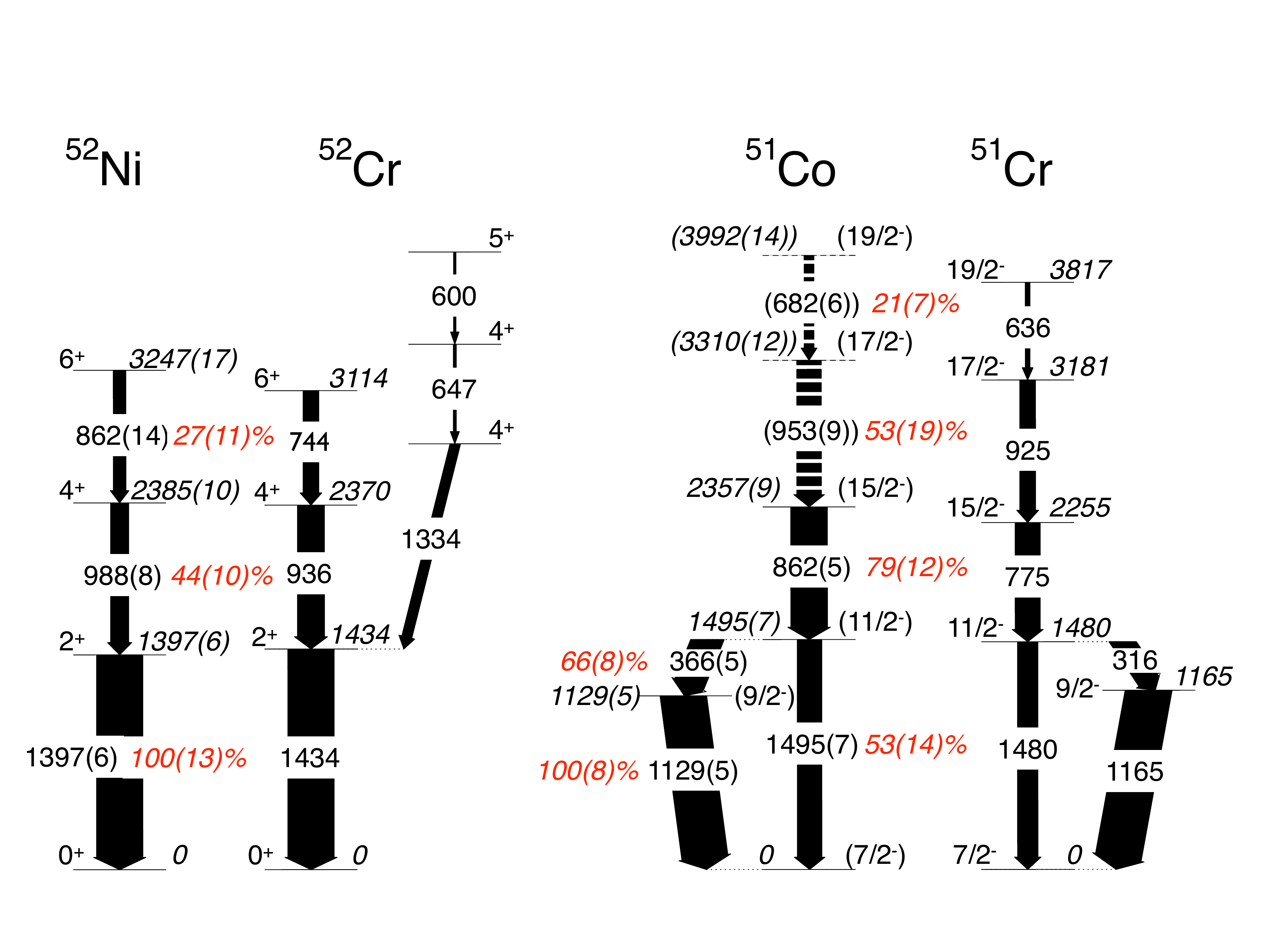}
\caption{The energy level schemes for \nuc{52}{Ni}~and \nuc{51}{Co} deduced from the current work. Tentative assignments are indicated by dashed lines, and the efficiency corrected relative intensity of $\gamma$-rays in \nuc{52}{Ni} and \nuc{51}{Co} are given in italic red text as a percentage. These are compared with the partial level schemes of \nuc{52}{Cr} and \nuc{51}{Cr}.}
\label{schemes}
\end{figure}

Figure \ref{spectra}(a) shows the Doppler-reconstructed spectrum for $\gamma$-rays found in coincidence with \nuc{52}{Ni}~ fragments in the S800. The spectrum is clearly dominated by three $\gamma$ rays, which are assigned as transitions from the \tpl, \fpl, and \spl~states. The resulting level scheme of \nuc{52}{Ni}~is shown in figure \ref{schemes}, which also shows a partial level scheme of \nuc{52}{Cr}~using information from \cite{Huo:2007wm}, but only showing transitions observed in the current work. The assignment of these transitions to the yrast sequence of \nuc{52}{Ni} is based on (a) the intensity profile of the $\gamma$ rays (decreasing intensity with increasing spin) and (b) mirror symmetry arguments -- i.e. through comparison with the spectrum of \nuc{52}{Cr}, presented in figure \ref{spectra}(b), populated by the mirrored reaction. Additionally, confidence is given to these assignments through comparison with two-neutron cross-section calculations, see the later discussion. 

The Doppler corrected $\gamma$-ray spectrum for \nuc{51}{Co}, populated by two neutron knockout from the \nuc{53}{Co} beam, is presented in figure \ref{spectra}(c). Six $\gamma$-ray transitions are observed, and the comparison with the spectrum in figure \ref{spectra}(d) for \nuc{51}{Cr} indicates, at least initially, the one-to-one correspondence between the mirrored transitions as shown by the dashed lines. The known partial level scheme of \nuc{51}{Cr} is shown in figure \ref{schemes} \cite{Xiaolong:2006ti}, and the proposed scheme for \nuc{51}{Co}, based on the spectral comparison, is also presented in figure \ref{schemes}. However, care must be taken in making these initial assignments, as the population mechanisms for the spectra in figures \ref{spectra}(c) and (d) are rather different, and the states of interest in \nuc{51}{Co} are well above the proton-separation energy. To give more confidence to the proposed scheme, two-neutron knockout cross-section calculations were utilised.

Two-neutron cross sections were calculated using the formalism presented in Refs. \cite{Simpson:2009p132502,Tostevin:2006ev}, combining eikonal reaction dynamics and shell-model structure input. The two nucleons are assumed to be suddenly removed from the projectile, the core (i.e. reaction residue) of which acts as a spectator during the reaction. Full $fp$-shell-model calculations using the KB3G interaction \cite{Poves:2001fi} were used to compute the two nucleon amplitudes (TNA), the amplitudes for each two-nucleon configuration with angular momentum $J^\pi$, coupled to residue state $J_f^\pi$ in the projectile ground state $J_i^\pi$. Valence nucleon radial wave functions are calculated in a Woods-Saxon plus spin-orbit potential, the geometry of which is constrained by Hartree-Fock (HF) calculations. For convenience of the interface with the reaction model, these structure amplitudes were computed using NuShell@MSU \cite{Brown:tw,Brown:1998fx}.

\begin{figure}
\includegraphics[width=\columnwidth]{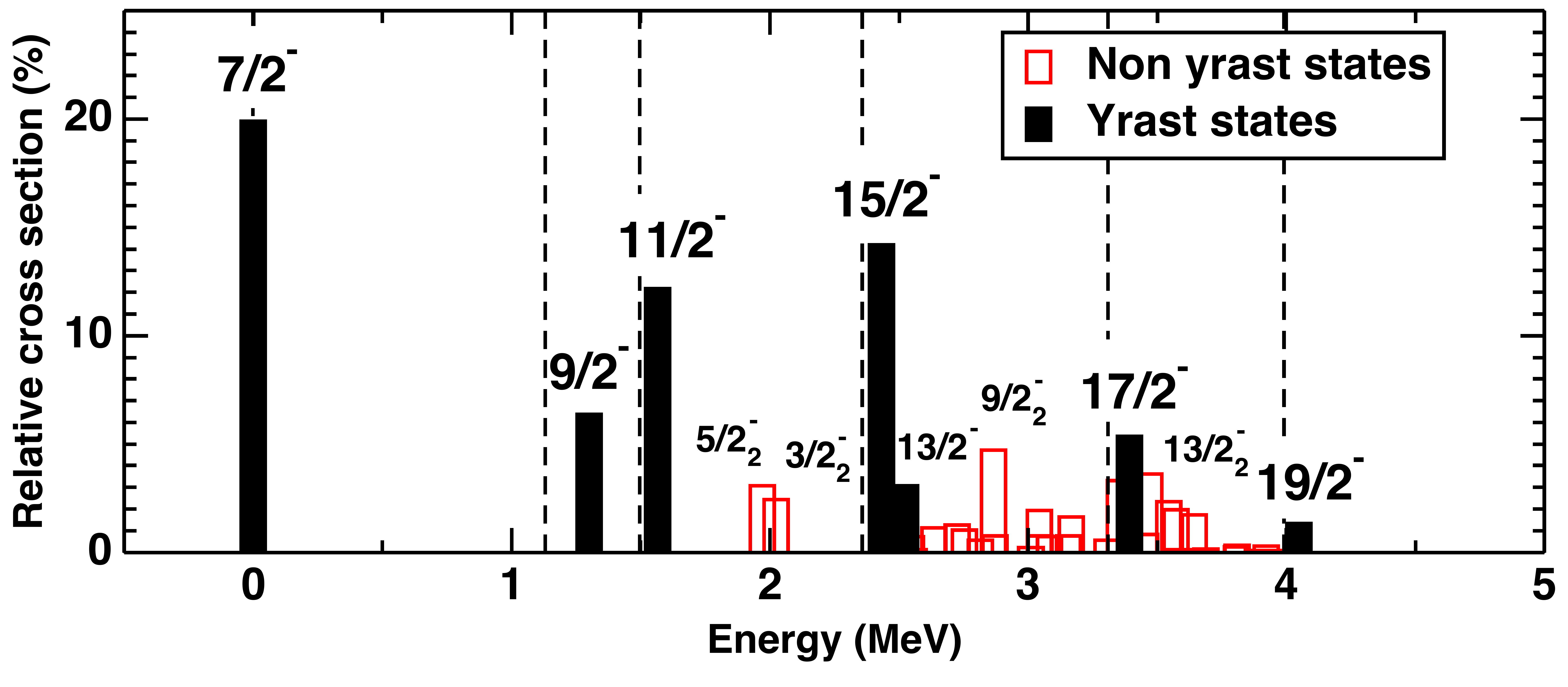}
\caption{Calculated relative cross sections for states in \nuc{51}{Co} populated in a direct two-neutron knockout reaction, using shell model input from NuShell@MSU with the KB3G interaction (see \cite{Simpson:2009p132502,Tostevin:2006ev}~ for details). The fraction of the intensity directly feeding the yrast (black) and non-yrast (red) states is shown. The horizontal axis is the calculated energy of the states from the shell model, and the dashed lines indicate the observed energies of the states in this work.}
\label{branches}
\end{figure}

For \nuc{52}Ni the calculations predict that 75$\%$ of the cross-section for excited states goes directly to the yrast \tpl, \fpl~and \spl~states, and predicts that the $\gamma$-ray from the \spl~state will be at least twice as intense as any from the non-yrast states. This is entirely consistent with the intensities in the observed spectrum and lends weight to the assignments presented in figure \ref{schemes}. For \nuc{51}{Co} the situation is more complicated as the intensity is expected to be more fragmented - see figure \ref{branches} for the calculated relative cross sections. It is clear from the spectra and $\gamma$-ray energies, that the 366, 1129 and 1495 keV transitions must form decays from the $\frac{9}{2}^-$ and $\frac{11}{2}^-$ yrast states in \nuc{51}{Co}. The calculations confirm that these two states should be strongly populated. In addition, the calculations show that we expect to see a strong transition from the \fthm state. Of the remaining (unassigned) transitions in figure \ref{spectra} (c), only the 862 keV transition is a candidate for this transition, from intensity arguments. This, along with the obvious mirror symmetry, is sufficient to confirm this assignment. For the remaining two transitions, at 953 and 682 keV, it is tempting to assign these to the next two members of the yrast sequence, the \sthm and \nthm states, based on mirror-symmetry arguments and measured intensities. However, the predicted relative cross sections in figure \ref{branches} do not indicate a strong population of these states and so these two assignments remain tentative in figure \ref{schemes}. For all states identified in this work, the spins and parities are inferred, rather than directly measured, and so they are indicated in figure~\ref{schemes} as tentative.



\begin{figure}
\includegraphics[width=\columnwidth]{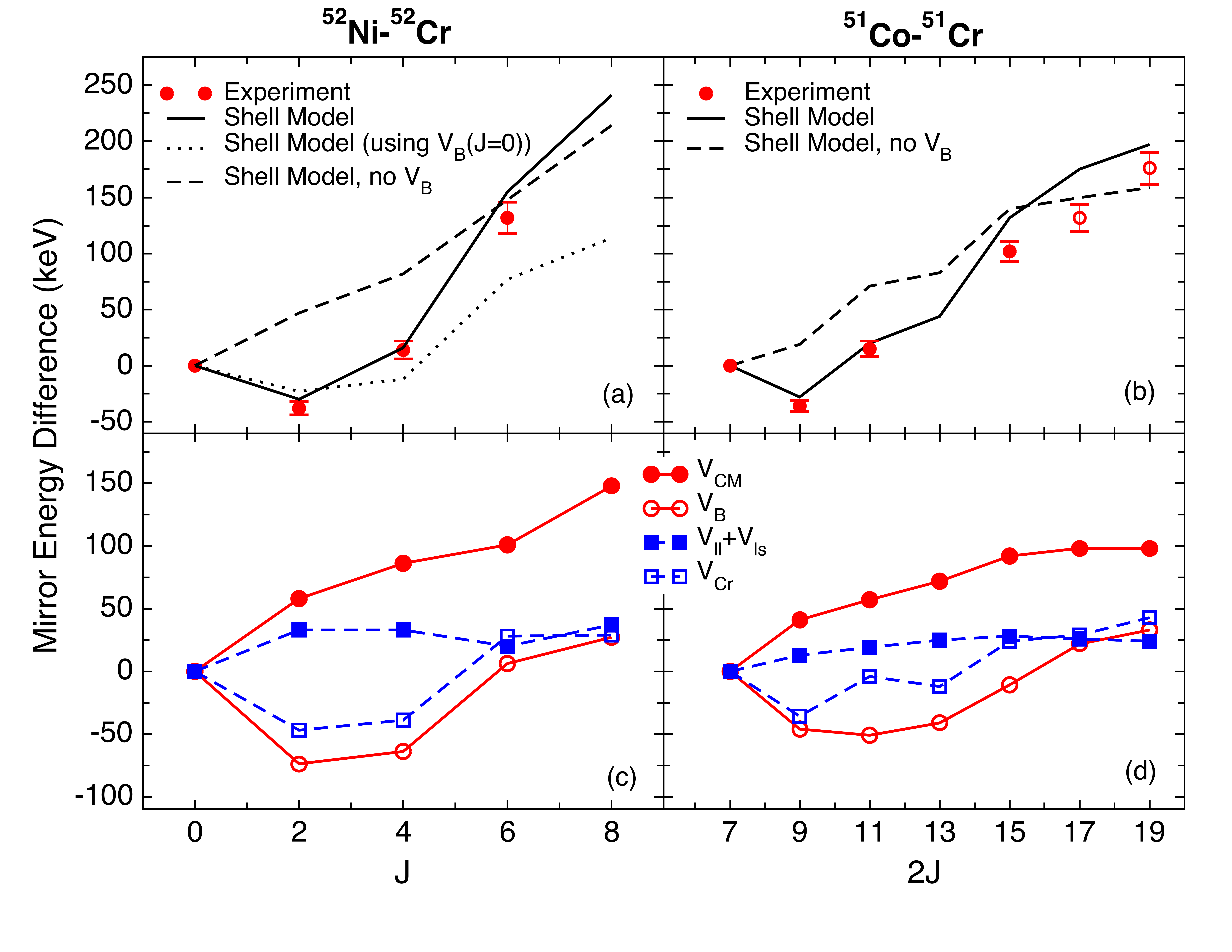}
\caption{(a) and (b): Comparison of the experimental MEDs with the predictions from shell-model calculations (solid line), using the formalism described in the text which includes the V$_{B}$ term (for $J=2$). Calculations without the inclusion of V$_{B}$ are shown by the dashed line. Additionally, in (a), the dotted line shows calculations where the usual V$_{B}$ term (for $J=2$) is replaced with a term of opposite sign for $J=0$ couplings -- see text for details. (c) and (d):  The four isospin-breaking components of the shell-model calculation (see text) the total of which sum to give the solid line in (a) and (b)}
\label{MED}
\end{figure}

The resulting MED for the \afiftytwo~ and \afiftyone~ mirror pairs are shown in figure \ref{MED}(a) and (b), respectively (calculated as: $MED(J) = E^{*}_{J,T,-T_z}-E^{*}_{J,T,T_z}$, where $E^{*}_{J,T,T_z}$ is the excitation energy of a state with spin $J$, isospin $T$ and isospin projection $T_z$). The observed rise in the MED is easily explained \cite{Bentley:2007fe} as due to the recoupling of angular momentum with increasing spin for neutron pairs in \nuc{52}{Ni} and protons pairs in \nuc{52}{Cr}. Such recoupling reduces the overlap of nucleons and hence, for protons, also the Coulomb energy. This yields a positive MED, typically of around 100 keV for nucleons in the \fsh~shell \cite{Bentley:2007fe}. For the $A=51$ mirror pair, which can be viewed as a proton(neutron)-hole in \nuc{52}{Ni}(\nuc{52}{Cr}), we see a similar trend in the MED, figure \ref{MED}(b), suggesting that the same structural effects are occurring. 

Large-scale shell-model calculations using the full $fp$-shell have been performed using the Antoine code \cite{Caurier:2005hc} with the KB3G \cite{Poves:2001fi} interaction. Previous work \cite{Poves:2001fi} shows that in this upper part of the \fsh~shell, and in the $A=51$ and 52 isobars in particular, the agreement with data on excitation energies and transition strengths is excellent. In order to calculate the isospin-breaking effects and their contribution to the MED, an identical approach to that described in \cite{Bentley:2007fe} has been employed. The model has four components that contribute to the MED: (a) The Coulomb multipole effect ($V_{CM}$) accounts for the recoupling effect described above, and is accounted for by the addition of the Coulomb energy to the two-body matrix elements for protons; (b) The radial term ($V_{Cr}$) is a monopole term that accounts for the Coulomb energy of changes in nuclear radius, according to the prescription of \cite{Zuker:2002cf}; (c) The $V_{ll}$ and $V_{ls}$ terms are monopole terms that account for Coulomb \cite{Zuker:2002cf} and magnetic \cite{Nolen:td} shifts to single-particle levels; (d) The final term ($V_B$) accounts for the $J=2$ effect described earlier - which is found to be necessary for a successful MED description in the shell \cite{Bentley:1998dh, Bentley:2007fe,Williams:2003km,Zuker:2002cf}. It is included in the model by adding a single repulsive interaction of 100 keV to the proton two-body matrix elements for \fsh~protons.  This shell-model prescription has been shown to be extremely successful in describing the $J-$dependence of MED for $T=\frac{1}{2}$ and $1$ states, but has yet to be tested in detail for such a large difference in $T_z$, where the monopole terms (which scale with difference in proton number)  will become large. In addition, for states far above the proton separation energy (i.e. the excited states in \nuc{51}{Co}) the effects of coupling to the continuum may become more important \cite{Michel:2010ke}.

The calculated and experimental MED are compared in figure \ref{MED}(a) for \afiftytwo~and (b) for \afiftyone. The solid line shows the prediction of the MED including all four effects as described above, and each of the four individual components to the MED are plotted in figure \ref{MED}(c) and (d). A number of points become apparent in this comparison. Firstly, the overall agreement is extremely good for both sets of nuclei, the greatest deviation between calculation and data being the tentatively assigned highest spin states in \nuc{51}{Co}. Secondly, it can be seen that the two monopole components $V_{ll}$/$V_{ls}$ and $V_{Cr}$ (dashed lines in figures \ref{MED}(c) and (d)), which scale with difference in $Z$, are  large -- and comparable with the multipole terms at low spin. Finally, by comparing the two multipole terms it is clear that at low spin $V_B$ (open circles), which represents the $J=2$ anomaly, is of comparable magnitude to the Coulomb multipole term, $V_{CM}$ (closed circles), and of opposite sign. This has the effect of virtually cancelling $V_{CM}$ for both the \tpl~ and \fpl~ states.

The clear need for the additional $J=2$ isovector interaction is demonstrated in figure \ref{MED}(a), where the dashed line shows the effect of not including the $V_B$ term. $V_B$ will contribute strongly wherever there is a significant change in the $T=1, J=2$ component of the wave function for protons in {\sl one} nucleus (and hence neutrons in its mirror). An example is the mirror-pair \nuc{54}{Ni}/\nuc{54}{Fe}~\cite{Gadea:2006gg,Rudolph:2008et} (expected to have a \fsh$^{-2}$ structure) where $V_B$ was found to be significant for the \tpl~ state. In the case of \nuc{52}{Ni}(\nuc{52}{Cr}), with four neutron(proton) holes in doubly-magic \nuc{56}{Ni}, the effect is amplified by the bigger difference in proton number, and the wave functions of both the \tpl~ and \fpl~ states are expected to have significant $J=2$ components. This is shown very clearly by the open circles in figure \ref{MED} (b), and figure \ref{MED} (a) provides the most convincing evidence to date that the anomalous $J=2$ isovector component of the two-body interaction for the \fsh~shell must be used to obtain agreement between the shell model and experiment. For the \afiftyone~case,  the $J=2$ strength is distributed among a wider range of states, and so the effect is less clear. Nevertheless, as figure \ref{MED}(b) shows, the need for $V_B$ in this case is equally compelling.

Since these MED are normalized to the ground state, the analysis presented here is sensitive to the $J-$dependence of these effects. Thus, one might expect that rather than a repulsive $J=2$ component being added (for protons) one could include an attractive $J=0$ component instead. This was first investigated in reference~\cite{Rudolph:2008et}, which showed that the $J=0$ approach did not have the required effect at higher spins. This is confirmed in figure \ref{MED}(a) where the dotted line shows the effect of including an attractive $J=0$ component for protons. This obviously fails to follow the data for the \spl~ state. It is worth stressing that this region of nuclei is very well-described by large-scale shell-model calculations in the \fp~ valence space, and yet the evidence for the need for inclusion of this additional isovector term within this shell-model description is overwhelming.

It is timely to consider possible origins. If the anomaly can be genuinely assocoiated with a multipole phenomenon, as seems to be likely, then the source can only be Coulomb or nuclear (i.e. charge-symmetry breaking, CSB) in origin. The observed spin-depedence of the additional effect (increasing from $J=0$ to $J=2$) rules out a simple two-body Coulomb term alone, as the wave-function overlap of the pair must {\it reduce} with increasing $J$. Of course, we cannot rule out other Coulomb-induced effects, and attempts have been made~\cite{Lenzi01} at a renormalisation of the Coulomb matrix elements to account for these, but the required $J$-dependence could not be reproduced. In terms of CSB, it has long been recognised that accurate calculations of nuclear masses (e.g. to predict location of the proton drip line) requires inclusion of CSB effects in order to correctly determine the Coulomb Displacement Energy (CDE). However, the numerical values needed are not known and usually {\sl effective} isovector matrix elements have been determined by fitting experimental CDE (e.g.~\cite{ormand89}). A recent study from Kaneko et al.~\cite{kaneko13}, parallel to the work presented here, is an example of this. In that work, an isospin non-conserving isovector term of 100 keV (attractive for protons in the $J=0$ channel) was used as it was found necessary in order to reproduce the detailed behaviour of the CDE. Thus, there are some indications that the two observations -- one concerning ground state energies, the other MED -- may have the same origin in this region. However, associating these effects with the nuclear interaction is possibly not safe at this point, and clearly, some theoretical effort is needed to understand the origin from both a nuclear structure, and fundamental interaction, perspective.





This work was supported by the UK Science and Technology Facilities Council (STFC) under grant numbers ST/J000124, J000051, and J000132 and the National Science Foundation (NSF) under grant number PHY-0606007.


%

\end{document}